**Ultrafast scanning electron microscope applied for studying the interaction between free electrons and optical near-fields of periodic nanostructures**

M. Kozák[1,2,*], J. McNeur[1], N. Schönenberger[1], J. Illmer[1], A. Li[1], A. Tafel[1], P. Yousefi[1], T. Eckstein[1], and P. Hommelhoff[1]

[1] *Department of Physics, Friedrich-Alexander-Universität Erlangen-Nürnberg (FAU), Staudtstrasse 1, 91058 Erlangen, Germany, EU*

[2] *Faculty of Mathematics and Physics, Charles University, Ke Karlovu 3, 12116 Prague 2, Czech Republic, EU*

[*]Corresponding author: martin.kozak@fau.de

**Abstract:**

In this paper we describe an ultrafast scanning electron microscope setup developed for the research of inelastic scattering of electrons at optical near-fields of periodic dielectric nanostructures. Electron emission from the Schottky cathode is controlled by ultraviolet femtosecond laser pulses. The electron pulse duration at the interaction site is characterized via cross-correlation of the electrons with an infrared laser pulse that excites a synchronous periodic near-field on the surface of a silicon nanostructure. The lower limit of 410 fs is found in the regime of a single electron per pulse. The role of pulse broadening due to Coulomb interaction in multielectron pulses is investigated. The setup is used to demonstrate an increase of the interaction distance between the electrons and the optical near-fields by introducing a pulse-front-tilt to the infrared laser beam. Further we show the dependence of the final electron spectra on the resonance condition between the phase velocity of the optical near-field and the electron propagation velocity. The resonance is controlled by adjusting the initial electron energy/velocity and by introducing a linear chirp to the structure period allowing to increase the final electron energy gain up to a demonstrated 3.8 keV.

## I. INTRODUCTION

Inelastic scattering of electrons by optical near-fields, excited by femtosecond laser pulses in the vicinity of various nanoobjects, has been studied in recent years from different perspectives. The resulting energy modulation imprinted to the electron beam on sub-optical cycle time scales is interesting for various fields of physics. It is considered for electron acceleration [1-13], for enhancing the visibility of low-contrast nanostructures in photon-induced near-field electron microscopy (PINEM) [14-17], for studying the quantized interactions between light and electrons [18-22] as well as for improving the temporal resolution of ultrafast electron diffraction and microscopy experiments [20,23-29]. This technique can help to overcome the temporal resolution limitations (typically few hundreds of femtoseconds) given by the dispersive broadening of electron pulses during their propagation from the source to the specimen. Achieving sub-optical cycle temporal control of freely propagating electrons by their coherent interaction with light may enable direct access to probing ultrafast coherent electronic dynamics with electrons or the full characterization of optical near-fields of various nanostructures, including phase-resolved spectroscopy.

The interaction between electrons and optical near-fields is based on modifying the dispersion relation of light propagating in vacuum close to an object with refractive index $n>1$. The phase velocity of the evanescent near-field can be matched to the propagation velocity of an electron near the scattering object, leading to a synchronous interaction between the field and the electron [6-10,15,16,30]. The spatial distribution of the electromagnetic near-fields in the vicinity of a nanostructure can be described using, e.g., Mie scattering theory in the case of a single nanosphere [16] or numerical techniques in the case of more complex nanostructures of various shapes [9,30]. Generally, the near-field amplitude decreases with increasing distance from the object ($\sim 1/r^3$ for a sphere, $\sim e^{-r/\Gamma}$ for periodic structures) on sub-wavelength scales. Hence to maximize the current of electrons interacting with the generated evanescent field, the transverse dimensions of the electron beam have to be smaller than the field decay length (typically $\Gamma=\lambda\beta\gamma/2\pi$=10-100 nm, where $\lambda$ is the light wavelength, $\beta$ is electron velocity in units of speed of light $c$ and $\gamma=\left(1-\beta^2\right)^{-1/2}$ is the Lorentz factor of electrons). Further, due to the necessity of high field amplitudes allowing reaching measurable electron energy modulation, short laser pulses with durations of fs-ps (∼$10^{-15}$-$10^{-12}$ s) are required for the excitation of the near-fields. For the interaction of all generated electrons with the optical near-fields, the duration of the electron pulse has to be comparable to or shorter than the laser pulse duration.

These two requirements, namely the electron beam that can be focused to a spot with clearly sub-micron-sized transverse dimensions and a pulsed operation with femtosecond duration of electron pulses, are met in ultrafast electron microscopes, where the electron emission is triggered by ultrashort laser pulses [31-38]. There are several reasons why transmission electron microscopes (TEMs) equipped with field-emission electron sources are considered as ideal for this application. They offer a high degree of transverse coherence (coherence length of ∼1 μm [37]) and monochromaticity (absolute energy spread of $\Delta E$~0.5 eV, resulting in a relative energy spread of $\Delta E/E=10^{-5}$-$10^{-6}$) of the electron beam. Further advantages of TEMs are built-in high quality imaging systems and the possibility to acquire images and diffraction patterns in the same measurement setup by adapting the electron imaging optics. The energy resolution of spectrometers used for electron energy-loss spectroscopy (EELS) of ∼0.1 eV further enables spectroscopy of the electrons after their interaction with optical fields and allows resolving quantum coherent features [14-16,18-21]. However, for some applications, the energy acceptance window of these spectrometers, which is typically limited to ∼10 keV, is not sufficient. In experiments focused on electron acceleration by laser fields, the observed energy gains at sub-relativistic electron energies are already approaching several keV [9,10] and higher gains are expected in the future [8]. Furthermore, the specimen chamber in most TEMs has very small dimensions (few millimeters in the electron beam direction). This fact significantly limits the freedom of choice of light coupling geometries and also makes the implementation of experiments with more than one laser beam complicated.

In this paper we describe the development of an experimental setup based on an ultrafast scanning electron microscope (USEM) [38] equipped with a heated Schottky-type field-emission tip cathode (in the following referred to as Schottky cathode). The setup serves for the research of the inelastic interaction between free electrons and optical near-fields and will be used in the future as a tool for studying different structure geometries and coupling schemes for efficient electron acceleration [9,39-41] and for the implementation of different techniques for transverse and longitudinal manipulation with freely propagating electrons on sub-optical cycle time scales in time-resolved electron imaging and diffraction experiments [23,29,42]. The vacuum chamber of the USEM accommodates both a dielectric nanostructure, where the optical near-fields are generated by femtosecond laser pulses, and a detection setup based on an electromagnetic spectrometer and a microchannel plate (MCP) detector, which allows us to measure the post-interaction electron spectra. The spectra are studied as a function of the time-delay between the pulsed electron beam and the pulsed optical near-fields, the electron initial energy and/or the parameters of the nanostructure (material, geometry, etc.). This paper is focused on describing the details and capabilities of this new USEM. Furthermore, we show a few examples of applications of this setup. We investigate the temporal broadening of the electron pulses due to repulsive Coulomb interaction between the electrons. Further we study the role of the resonance condition between the electrons and the synchronous optical near-field mode. Finally we demonstrate the extension of the interaction length in electron acceleration driven by pulse-front-tilted laser beam at a chirped grating nanostructure.

## II. EXPERIMENTAL SETUP

### A. Laser source

In the experiment, two synchronized femtosecond laser pulses are used both to temporally control the emission of the electron pulse and to excite the optical near-fields in the interaction region inside the USEM vacuum chamber (see the layout of the experimental setup in Fig. 1). Ultraviolet (UV) laser pulses induce electron photoemission in the USEM electron gun while infrared (IR) laser pulses excite the optical near-fields on the surface of the nanostructure. The IR pulses are generated in an optical parametric amplifier (OPA) pumped by a Ti:sapphire regenerative amplifier running at a repetition rate of $f_{\mathrm{rep}}$=1 kHz. The small repetition rate was chosen to achieve high peak powers, which enables increasing the interaction distance between electrons and laser fields in acceleration experiments while keeping the amplitude of the field strength in the order of 10 GV/m. Depending on the wavelength, the pulse duration and the pulse energy of the OPA signal and idler waves are $\tau_{\mathrm{FWHM}}$=50-100 fs and $E_{\mathrm{p}}$>100 μJ. The UV pulses for electron photoemission are generated by sum-frequency mixing the signal wave from the OPA (wavelength of 1100-1600 nm) and part of the basic amplifier output at $\lambda_0$=800 nm, and subsequent second harmonic generation. The UV laser beam at a wavelength of $\lambda_{UV}$=251-285 nm is focused on the USEM Schottky cathode through the vacuum window from a direction perpendicular to the tip symmetry axis with linear polarization parallel to the electron emission direction. The pulse energy is 0.5-30 nJ and the spot radius is $w_{UV}\approx$6 μm. The laser beam is aligned to the front facet of the Schottky cathode by heating the tip to high temperature ($T$>1500 K). The black-body radiation in the visible and infrared is used to align two irises in the beam-path by maximizing the power transmitted through the two apertures using a CCD camera. The UV laser beam is then aligned to these two irises. The fine alignment is done by optimizing the emitted electron current detected by the MCP.

**B. Pulsed electron beam**

The femtosecond pulsed electron beam is generated by photoemission in a standard SEM (FEI XL 30 FEG) equipped with a Schottky cathode [43]. The cathode consists of a tungsten tip with a flat front facet (100-500 nm in diameter) oriented in <100> crystallographic direction. The electrons are emitted by a single-photon process using the UV femtosecond pulses to control the emission time. The height of the surface potential barrier for electrons at the front facet of the cathode (workfunction at <100> tungsten surface is $\Phi$=4.6 eV [44]) is lowered to 2.8-3 eV by the Schottky effect due to the applied field strength of 0.8 GV/m and by a layer of $ZrO_x$ which is supplied on the tip front facet when the tip is heated to 1800 K during continuous operation [43]. However, when the tip is kept at room temperature to suppress the DC electron current in the photoemission operation mode, the barrier height is slowly increasing on time scales of several hours (see Figure 1(b)), probably due to deposition of impurities and adsorbates on the surface and slow removal of $ZrO_x$ by laser illumination. This drop of the electron current is present even for relatively good gun vacuum levels of $p<4\times10^{-10}$ mbar. We found that the increase of the effective barrier height saturates at the value of $\Phi_{eff}$=4.5-5 eV determined from the photon energy, at which the single-photon photo-emitted current drops to zero after approx. 1 hour of operation. A similar growth of the effective barrier height was previously observed in experiments with optical field emission from tungsten nanotips [34]. To reach single-photon operation with a stable emission current, we typically use the wavelengths of $\lambda_{UV}$=251-285 nm corresponding to the photon energy of $E_{UV}$=4.35-4.95 eV.

The decrease of the electron current with time can be partially suppressed by heating the cathode to a moderate temperature of 1100-1300 K, where the DC current is still negligible but the contamination of the surface is slower due to the elevated temperature. However, this temperature level is still too low to reach diffusion of $ZrO_x$ from the reservoir to the tip apex. Due to this reason we regularly flash the tip by heating it to temperatures of ~1600 K (similar to [37]). After this procedure, the photoemission current returns back to its original value.

To find out the influence of the laser illumination on the drop of the photoemission current we measured the current with different time delays after flashing the tip with and without continuous illumination by fs pulses. Here we observed that the trend of decreasing emission current with time is independent of the illumination but the slope of this decrease $-dI/dt$ growth with laser power. From the results we anticipate that the illumination plays a significant role only at high power levels. There are two possible reasons for this behavior: 1) The multiphoton ionization of residual gas in the gun leads to production of ions that are attracted by the cathode. When considering the gun pressure, laser repetition rate and illuminated volume, enough ions can be produced to cover tens of percent of the front facet of the tip after one hour. 2) At high intensities, also the surface geometry of the front facet of the Schottky cathode can change, similar to observations made with a cold field-emission tip [45]. This influences the field enhancement on the surface and thus both the emitted electron current and the spatial distribution of emitted electrons.

After photoemission, the electrons are accelerated by electrostatic fields to the final kinetic energy of $E_{kin}$=1-30 keV. The electron beam is focused by the objective lens to the focal plane with a working distance of $w_d$=20 mm reaching a transverse spot size of $w_e$~50-100 nm ($1/e^2$ radius) using a 100 μm diameter objective lens aperture (the objective aperture is shown in Fig. 1). To increase the number of electrons available for the interaction, the condenser lens of the USEM column is set to the highest probe current setting, in which the electron beam is almost perfectly collimated during its propagation through the column (see the calculated electron trajectories in Fig. 5).

**C. Optical near-fields, pulse-front-tilt**

The optical near-fields used for the interaction with electrons are excited by the IR pulses on the surface of a nonresonant silicon nanograting [10,23]. The spatial distribution of the synchronous spatial harmonics can be described as an evanescent wave exponentially decaying with the distance from the surface and propagating in the direction of the grating *k*-vector [13]. The cycle-averaged force acting on the electrons during the interaction in the case of nonresonant structures (without resonant enhancement of the near-field amplitude of the synchronous mode) can be written as $F(t,z) \simeq |E(t)| \exp(-\Gamma x) \sin(\omega t)$, where $E(t)$ is the temporal envelope of the driving laser pulse electric field.

The time delay between the UV and IR laser pulses is controlled by a standard optical delay line (0-600 ps, precision of 10 fs), effectively controlling the arrival time of electrons with respect to the optical near-fields. For the light coupling geometry with the laser and electron beams perpendicular to each other [30], the distance over which the electrons interact with the near-fields is limited by the pulse duration for the flat intensity-front beam. The interaction distance can be significantly increased by using a pulse-front tilted (PFT) laser beam [46,47] (see the sketch of the interaction between the travelling electron and the PFT laser beam in Fig. 2 (a)) generated by a dispersive element and the imaging optics shown in Fig. 1 (a). While the intensity fronts are tilted, the phase-fronts of such a spatio-temporally modulated beam are still perpendicular to the propagation direction. Therefore the coupling to the evanescent near-field mode is not affected. In fact, the group velocity of the envelope of the synchronous near-field is matched to its phase velocity. By imaging the surface of the reflective diffraction grating using a cylindrical lens with focal distance of $f_{cyl}$=70 cm and the final focusing lens (asphere, $f_{fin}$=25 mm), an intesity-front angle $\theta_{PFT}$ is reached in the interaction region [48]. The magnification of the imaging setup in the dispersion plane determines the angle $\theta_{PFT}$=76°, which fulfills the relation $\tan(\theta_{PFT})=\tan(\theta) f_{cyl}/f_{fin}$, where $\theta$=8° is the diffraction angle of light at wavelength $\lambda$=1.93 µm at the diffraction grating. To avoid spatio-temporal distortions of the laser beam, we use a geometry in which the diffracted beam is perpendicular to the diffraction grating [49].

The PFT laser beam at a wavelength of $\lambda$=1.93 μm is characterized using cross-correlation with a flat intensity front beam on a silicon-based charge-coupled device (CCD) chip utilizing the two-photon absorption process of the IR pulse (see the sketch of the cross-correlation measurement in the inset of Fig. 2 (b)). The transverse position of the peak of the cross-correlation signal in the electron propagation direction $z$ is plotted in Fig. 2 (b) as a function of the longitudinal shift of the PFT beam with respect to the flat intensity-front beam. The measured PFT angle $\theta_{PFT}$=74° leads to perfect synchronization of the group velocity of the near-fields with electrons propagating along the grating surface with velocity $v_e=1/\tan(\theta_{PFT})c=0.29c$ ($c$ is speed of light). With the PFT laser beam, the interaction distance is only limited by the transverse laser spot size in the electron propagation direction (≈100 μm) and the electron beam dynamics during the interaction.

**D. Electron detection setup**

The setup for detection of electron energy spectra after the interaction with optical near-fields consists of a home-built Elbek-type electromagnetic spectrometer [42,50] and a microchannel-plate detector (MCP, Chevron type) with a phosphor screen imaged by a CCD camera (see the layout in Figure 3(a)). The spectrometer is designed to offer a large energy acceptance window of $\delta E_k$=20-80 keV and close to linear dispersion relation. Its performance is verified using a calibration procedure, where the initial electron energy is varied by changing the USEM DC accelerating voltage in the range 27.8-29.6 keV. The dispersion and resolution of the spectrometer are measured (see Figures 3(b)-(d)). The initial energy width of the electron distribution $\Delta E$∼0.5 eV is negligible for the resolution measurement. For data acquisition, two modes are available at different experimental conditions. At the low repetition rate of 1 kHz, electron counting mode is used to suppress the dark noise. Here the individual peaks in images from the CCD camera with amplitudes above a threshold, which is higher than the noise level, are attributed to single electrons located in the center of mass of each peak and integrated by an acquisition software. Via this procedure, the spectral resolution of <40 eV (see the measured FWHM of the response function in Figure 3(d)) and high signal/noise ratio are experimentally reached. This is close to the numerically calculated 20 eV limited by the pixel size of the MCP detector. In this regime, the electron current has to be low enough to allow resolving individual electron density peaks. The background in the measured spectra of 0.001 counts/(s.bin) is only caused by the dark count rate of the MCP. At higher electron currents, the second mode is used where the total above-threshold image intensity from the CCD camera is integrated. Here the spectral resolution is limited to ∼100 eV due to the spatial resolution of the MCP phosphor screen.

## III. NUMERICAL SIMULATIONS

A virtual model of all active elements within the electron gun head is built in order to solve for the electric field distribution in the gun head. The dimensions for this geometry as well as the applied voltages are taken from the technical drawings of the microscope itself. The Schottky cathode is modeled as a conical tip terminated by a hemispherical apex with a radius of curvature of $r$=470 nm. To mimic the end facet of the emitter, the cone tip is cut perpendicular to the cone axis such that a flat surface with 300 nm diameter is formed. The static electric fields shown in Fig. 4 are calculated using the electrostatics module of COMSOL multiphysics in two dimensions (all the elements are cyllindricaly symmetric). The field maps are then revolved around the electron beam axis to yield the fully three-dimensional field distribution.

Electron trajectories (see Fig. 5) are calculated by the 5-th order Runge-Kutta algorithm using General Particle Tracer (GPT). The initial electron distribution is defined in the following way. The particle coordinates are generated randomly over the end facet of the Schottky cathode, with a two dimensional Gaussian distribution with FWHM diameter of 300 nm. The initial energy distribution is:

$$f(E)=\frac{2\sqrt{\ln 2}}{\sqrt{\pi}\Delta E_0}e^{-4\ln 2\frac{(E-E_0)^2}{\Delta E_0^2}}, \tag{1}$$

where $E_0$ is the central energy and $\Delta E_0$ the FWHM energy width. The energy width defines the magnitude of the initial electron velocity. The direction of the velocity is uniformly distributed in the solid angle of $2\pi$ out of the tip surface. In the simulations, $E_0$=0.2 eV and $\Delta E_0$=0.5 eV (typical values for the single-photon photoemission from a Schottky tip [38]). The distribution of the electron emission time corresponds to the envelope of the UV laser pulse (Gaussian, $\tau_{FWHM}$=100 fs). We note that in addition to the electrostatic elements modelled in COMSOL, magnetic elements such as condenser lenses are included in the particle tracing simulations within GPT to accurately model trajectory effects during electron propagation through the microscope column. The Coulomb repulsion between the particles is taken into account to describe the space-charge effects on the final pulse duration and transverse dimensions of the probe beam. The position, velocity, energy and arrival time are evaluated at the interaction point of the experiment located 40 cm downstream from the tip apex.

Each simulation contains a set of $N$=3000 simulated electrons. For a correct description of statistical effects on the Coulomb interaction for few electrons per pulse, the number of electrons per pulse follows a Poisson distribution:

$$f\left(N, N_{\mathrm{av}}\right) = e^{-N_{\mathrm{av}}} \frac{\left(N_{\mathrm{av}}\right)^N}{N!}, \tag{2}$$

where $N_{\mathrm{av}}$ is the average number of electrons per pulse. The bunch duration is evaluated from the Gaussian fit of the histogram of arrival times of all simulated electrons to the interaction point.

## IV. RESULTS AND DISCUSSION

In this section we describe the experimental characterization of the femtosecond pulsed electron beam of the USEM. Further we show a few examples of measured electron spectra after inelastic scattering of electrons by optical near-fields of silicon nanostructures. To reach efficient energy transfer between the optical near-fields and the electrons, the resonance condition between the phase velocity of the $m$-th spatial harmonics of the near-field and the propagation velocity $\beta$ (in units of speed of light $c$) of the electrons has to be met [5,7,30,46]. This synchronicity condition is fulfilled if $\lambda_p = \beta\lambda m$, where $\lambda_p$ is the period of the structure. During the interaction, a time-periodic sinusoidal modulation of the electron energy is induced. The post-interaction electron energy spectra thus reveal information about the number/density of electrons present at the structure at the same time as the optical near-field pulse. By scanning the relative time delay between the optical pulses generating the near-fields and the electron pulses, the electron pulse duration is measured similar to refs. [19,23,51]. Another option for characterization of electron pulse duration is to use ponderomotive interaction between the electrons and light [52-55].

A set of parameters that fully characterize short electron pulses consist of the center energy $E_0$, an energy spread $\Delta E$ arising from the photoemission process and/or Coulomb interactions, the spatial coherence given by the transverse ($\varepsilon_\perp$) and the longitudinal ($\varepsilon_\parallel$) emittances of the beam, duration of the temporal envelope of the pulse and the total charge [56]. The temporal resolution of experiments with pulsed electron beams is limited by the achievable duration of the pulse envelope (individual pulse envelopes in the case of the attosecond pulse trains [20,26,29,42]) at the studied specimen. There are three main contributions to the final electron pulse duration. The first arises from the short times after photoemission, when the velocity of electrons is small and the relative velocity spread $\Delta v/v$ is high. In the case of flat cathodes, this leads to temporal broadening by $\tau_{acc} \approx \left(m_0 \Delta E/2\right)^{1/2} / eE_{acc}$, where $E_{\mathrm{acc}}$ is the homogeneous accelerating electrostatic field, $m_0$ the electron mass and $\Delta E$ the initial energy spread [57-60]. However, for tip-based electron sources, the field is strongly inhomogeneous along the electron trajectory. For this case, the acceleration contribution to the electron pulse broadening can be calculated numerically. The arrival time of the on-axis electrons as a function of the initial kinetic energy $E_{\mathrm{in}}$ (only longitudinal velocity component assumed) can be written as:

$$t = \sqrt{\frac{m_0}{2}} \int_0^d \frac{1}{\sqrt{q(U(0) - U(z)) + E_{\mathrm{in}}}} dz \, , \tag{3}$$

where $U(z)$ is the potential on the symmetry axis of the electron beam propagating along the $z$ direction. The temporal broadening of the pulse can be approximated as $\tau_{acc}=t(E_{\mathrm{in}}=0)-t(E_{\mathrm{in}}=\Delta E)$. The potential obtained from the numerical solution of the Laplace equation $\Delta U=0$ gives the field amplitude on the front facet of the Schottky cathode of 0.8 GV/m. This leads to a temporal broadening of $\tau_{\mathrm{acc}}\approx 30$ fs. The second contribution to the final electron pulse duration is given by the laser pulse duration of $\tau_{\mathrm{laser}} = 100$ fs. The final electron pulse duration (on-axis) can be calculated as $\tau_{electron} \approx \sqrt{\tau_{laser}^2 + \tau_{acc}^2}$. However, because of the relatively large emission angle of the electrons due to the distribution of the electrostatic field on the tip surface, additional broadening of the electron pulse arises from the trajectory effect (see the calculated electron trajectories in Fig. 5), the third contribution. The transverse distribution of electrons along the beam path leads to different lengths of their trajectories between the emission site and the interaction in the chamber. From numerical modelling we obtained a value of ∼400 fs for the minimum electron pulse duration, which can be obtained in the presented SEM-based setup.

### A. Electron pulse duration measurement

The electron pulse duration is measured by acquiring the post-interaction electron spectra as a function of the time delay between the UV laser photoemission pulse and the IR pulse (with PFT) that serves for the near-field generation on the surface of the single silicon grating with a period $\lambda_p$=620 nm. The electron beam energy in these experiments is 28 keV ($v_e=\beta c$=0.32 $c$) to fulfill the resonance condition. The electron velocity is thus slightly faster than the group velocity of the synchronous mode obtained using the PFT leading to shortening of the interaction distance to ∼50 μm. In Figure 6(a), the electron spectra are plotted as a function of the time delay between the UV and IR laser pulses. The cross-correlation signal (squares in Figure 6(b)) is obtained by integrating each electron spectrum out of the spectral window marked by the two dashed lines in Figure 6(a) (28.27-28.53 keV). Because the nanostructure used to generate the near-fields is nonresonant and because the tilted pulse fronts lead to equal values of the group velocity of both optical near-fields and electrons in the perpendicular coupling geometry, the temporal envelope of the near-fields in the electron rest frame is given by the temporal envelope of the laser pulse. For a sufficiently high energy cut-off and with laser pulses significantly shorter than the electron pulse duration, the cross-correlation signal corresponds to the temporal envelope of the electron pulse $f_e(t)$ [28].

The measured data is fitted with a Gaussian curve with $\tau_{e,FWHM}$=410±30 fs. This corresponds well to the numerical simulations predicting $\tau_{e,FWHM}$=400 fs. Such short pulses are only obtained in the regime of <1 electron/pulse emitted from the cathode. We further investigate the dependence of the electron pulse duration on the number of photoemitted electrons during a single laser pulse. For very high pulse charges corresponding to 500 electrons/pulse emitted from the cathode, the temporal envelope of the electron pulse shows a double-peak structure (see Figure 6(b)) due to the Coulomb repulsion shortly after emission. The two peaks correspond to the front and back part of the electron pulse that are repelled in the longitudinal direction and accelerated to the final energy. Such a double-peak structure is not observed for pulsed electron beams generated at flat photocathodes [61]. This can be explained by the difference in the current density shortly after photoemission. With 500 electrons per pulse emitted from the front facet of the tip, the maximum density is approximately 50 times higher than the current density with $10^4$ electrons per pulse in a typical experiment with a flat photocathode [61]. The Coulomb interaction causes acceleration of the front part and deceleration of the trailing part of the pulse, leading to the observed two peaks of electron density.

Apart from the initial energy spread due to the photoemission process, the electrons in the interaction region have a correlated energy spread due to the spectral broadening shortly after the photoemission and subsequent dispersive propagation. Therefore also the multi-electron pulses can be in principle compressed back to sub-picosecond durations by RF [62-64] or THz compression techniques [65].

In Figure 6(d) we compare the measured pulse duration as a function of electrons/pulse emitted from the cathode (lower scale) and delivered to the chamber (upper scale) with the numerical results. The probe current is limited by the diameter $d_{ap}$=100 μm of the USEM objective lens aperture and the setting of the condenser lens. The combination of these two parameters effectively sets the transverse emittance $\varepsilon_t$ of the probe beam, which needs to be lower than 0.1 nm.rad for experiments investigating the interaction between electrons and optical near-fields of periodic nanostructures due to the short transverse decay length of the near-fields. The probe current grows approximately linearly with the area of the objective lens aperture. However, for large aperture diameters, the transverse size of the probe beam in the focus grows significantly (up to ∼1 μm with $d_{ap}$=500 μm) due to aberrations of the electron optics. For experiments with less stringent requirements on the transverse emittance and spot size of the electron beam, currents corresponding to the lower scale in Fig. 6(d) can be used with the objective aperture fully open ($d_{ap}$>500 μm). Also the pulse duration grows with the aperture size due to the trajectory effects. The electrons propagating further from the symmetry axis have longer trajectories than the on-axis electrons leading to temporal broadening of the pulse. The minimum measured pulse duration with the objective lens aperture fully open is $\tau_{e,FWHM}$=730±30 fs [42].

**B. Electron spectra after the interaction with optical near-fields**

When the resonance condition is fulfilled, the electrons gain or lose kinetic energy dependent on their injection time with respect to the phase of the optical near-field and on their distance from the surface of the nanostructure [7,30]. The resulting electron spectrum is broadened with exponentially decaying tails (see the spectra with and without laser fields in Figure 7(a)). The exponential decay is a consequence of the transverse spatial shape of the accelerating/decelerating fields [30]. The energy gain of an electron interacting with a near-field mode with constant phase velocity in the impulse approximation assuming a small electron velocity change $\Delta v_e \ll v_e$ is given by the integral:

$$\Delta E = q \int_{-\infty}^{\infty} E_{\mathrm{long}}^{\mathrm{synch}}(\mathbf{r}, t) \mathrm{d}s \,, \tag{4}$$

where $q$ is the electron charge, $E_{\mathrm{long}}^{\mathrm{synch}}(\mathbf{r}, t)$ is the longitudinal component of the electric field of the synchronous harmonic and d$s$ is the element along the electron trajectory. This approximation, however, is only valid for our experimental parameters when $\Delta E$<0.5 keV. For higher amplitudes of the velocity modulation ($\Delta v_e/v_e$>1-3 %), the electrons slip over the initial phase of the near-field acting during the interaction and dephasing takes place [30]. This limits the maximum energy gain using periodic nanostructures to approx. 1-3 keV, depending on the interaction distance and the field amplitude applied [10].

**B.1. The role of the resonance condition for the interaction between electrons and near-field**

If the resonance condition is exactly fulfilled when the interaction starts and the electron velocity modulation during the interaction is small ($\Delta v_e/v_e$<0.5 %), the final electron spectra are symmetric as a consequence of the temporal periodicity of the net force applied to the electrons. This translates to the spatial periodicity in the electron propagation direction. The maximum energy gain and loss before the onset of dephasing is given by Eq. (4). However, when the electrons are initially slower than the optical mode, the spectrum becomes asymmetric due to different phase-matching (resonance) conditions for accelerated and decelerated electrons. In Figure 7(b,c) we show the measured (b) and calculated (c) spectra for electrons with different initial velocity $\beta$. The phase velocity of the accelerating mode was fixed in all the measurements. The deviation from resonance leads to a difference in the dephasing length and corresponding energy change of the electrons that gain or lose energy. If the electrons are initially slower than the optical mode, the acceleration brings part of the electron distribution to the resonance before dephasing occurs and these can be further accelerated to higher velocities. However, for the decelerated population, the dephasing comes earlier. As a consequence, the beam obtains a net energy gain.

**B.2. Controlling the resonance condition during the interaction**

To overcome the dephasing and to allow the electrons to be accelerated over longer distances, the phase velocity of the near-field mode has to be controlled during the interaction either by the frequency chirp introduced to the laser pulse or by the chirp of the period of the nanostructure. Because of the limited spectral bandwidth of the laser pulses we use the second approach. The structure period adiabatically grows along the electron trajectory as $\lambda_{\mathrm{p}} = \lambda_{\mathrm{p0}} + az$, where $\lambda_{\mathrm{p0}}$ is the initial structure period, $a \ll 1$ is the parameter of the linear chirp and $z$ is the electron propagation direction. In Figure 8(a) we show the electron spectra after the interaction with the optical near-fields of a chirped single-grating structure as a function of the chirp parameter $a$. The peak electric field of the laser beam with optimized PFT is measured to be 1.5 GV/m. The maximum measured energy gain as a function of $a$ is shown in Figure 8(b).

The highest observed energy gain of 3.8 keV is limited by several factors. The first limitation is due to the signal to noise ratio achievable in the presented setup with electron pulses containing less than 1 electron per pulse and the repetition rate of only 1 kHz. The second limiting factor is the sharp dependence of the longitudinal and transverse forces both on the injection phase of the electrons and on their distance from the surface of the nanostructure. Therefore the amount of electrons that can propagate along the chirped structure with the acceleration exactly matching the phase velocity of the mode decreases exponentially with the travelled distance. In other words, the electron beam experiences a non-uniform growth of both transverse and longitudinal emittance preventing further interaction with the appropriate phase of the optical near-field.

In RF accelerators, for instance, the electrons become relativistic during one field oscillation and then propagate with the velocity close to *c*. Therefore the relative changes of both transverse and longitudinal velocities during one period of the accelerating field becomes very small. However, even with GeV/m gradients reached in dielectric laser accelerators [66], the energy gain over one grating period for electrons initially at 30 keV is of the order of 1 keV. Therefore the electrons need to oscillate in the optical near-field many times before they reach relativistic energies. Further, the accelerating mode in a typical RF cavity is cylindrically symmetric allowing to use focusing elements along the beamline to keep the beam transversally confined. The lack of the transverse spatial symmetry of the accelerating fields makes the electron dynamics of dielectric laser accelerators extremely complicated. For the future success of the particle accelerators driven by optical near-fields, a configuration allowing stable acceleration of electrons trapped in a fraction of the phase space (both longitudinal and transverse) has to be developed.

## V. CONCLUSIONS

The experimental setup based on an ultrafast scanning electron microscope was developed for the investigation of the interaction between free electrons and optical near-fields or optical fields in general (see, for example, the inelastic scattering of electrons at a ponderomotive potential of an optical travelling wave described in [42]). Due to its variability it can also serve for time-resolved electron diffraction and microscopy experiments. The characterization of the pulsed electron beam confirms the possibility of directly (without further compression) reaching femtosecond temporal resolution with this setup with the lower limit for the electron pulse duration of $\tau_{e,FWHM}$=410 fs. The implementation of the pulse front tilt to the femtosecond pulsed laser beam further allows increasing the interaction distance between the electrons and high-amplitude optical near-fields far beyond the limit for the flat-intensity front beam. We show that by controlling the resonance condition between the propagating electrons and the optical near-fields, the shape of the electron spectra changes. By chirping the structure period, the synchronous interaction for electrons accelerated along the structure is reached leading to the increase of the maximum energy gain to 3.8 keV. Entering the regime of laser-driven electron dynamics brings many challenges for the further development of the dielectric laser accelerators. Reaching transverse and longitudinal beam stability will require the generation of focusing forces based on near-field [10] or ponderomotive interaction [67]. The implementation of these advanced schemes is the next step towards developing a miniaturized electron accelerator driven by optical fields. The USEM presented here will be helpful to provide the well-controlled electron beam with widely varying and well-matching beam parameters.

**ACKNOWLEDGMENTS**

We thank H. Deng, K. J. Leedle and J. S. Harris for fabrication a part of the nanostructures used in this study. The authors acknowledge funding from ERC grant "Near Field Atto", the Gordon and Betty Moore Foundation through Grant GBMF4744 "Accelerator on a Chip International Program – ACHIP" and BMBF via a project with contract number 05K16WEC.

Figures:

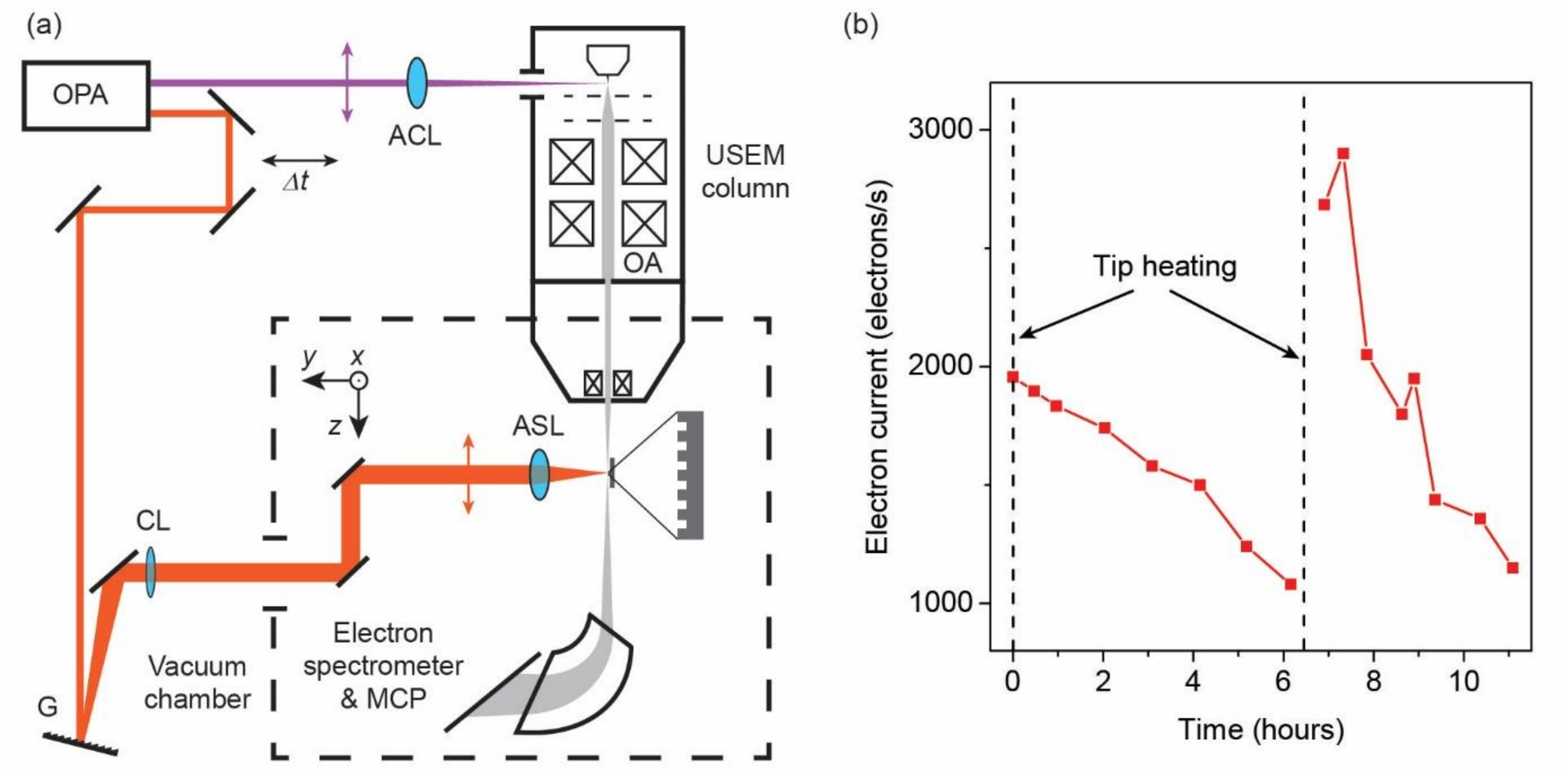

Figure 1. (a) Layout of the ultrafast scanning electron microscope (USEM) experimental setup. The pulsed UV laser beam (violet) is focused by an achromatic lens (ACL) to the USEM Schottky tip, where the electrons are photoemitted. The electron beam (grey) passes through the objective aperture (OA) and is focused to the interaction region close to the surface of a periodic dielectric nanostructure. The pulsed IR laser beam (red), which is used for optical near-field generation, is delayed by an optical delay line ($\Delta t$) and dispersed by a diffraction grating (G), whose surface is imaged by a cylindrical lens (CL) and an aspherical lens (ASL) to the surface of the nanostructure in the USEM vacuum chamber. Electron spectra are measured by an electromagnetic spectrometer and a micro-channel plate detector (MCP). (b) Photoemission electron current as a function of time from tip flashing (heating to ~1600 K). Excitation photon energy is $E_{UV}$=4.7 eV, pulse energy is 12 nJ.

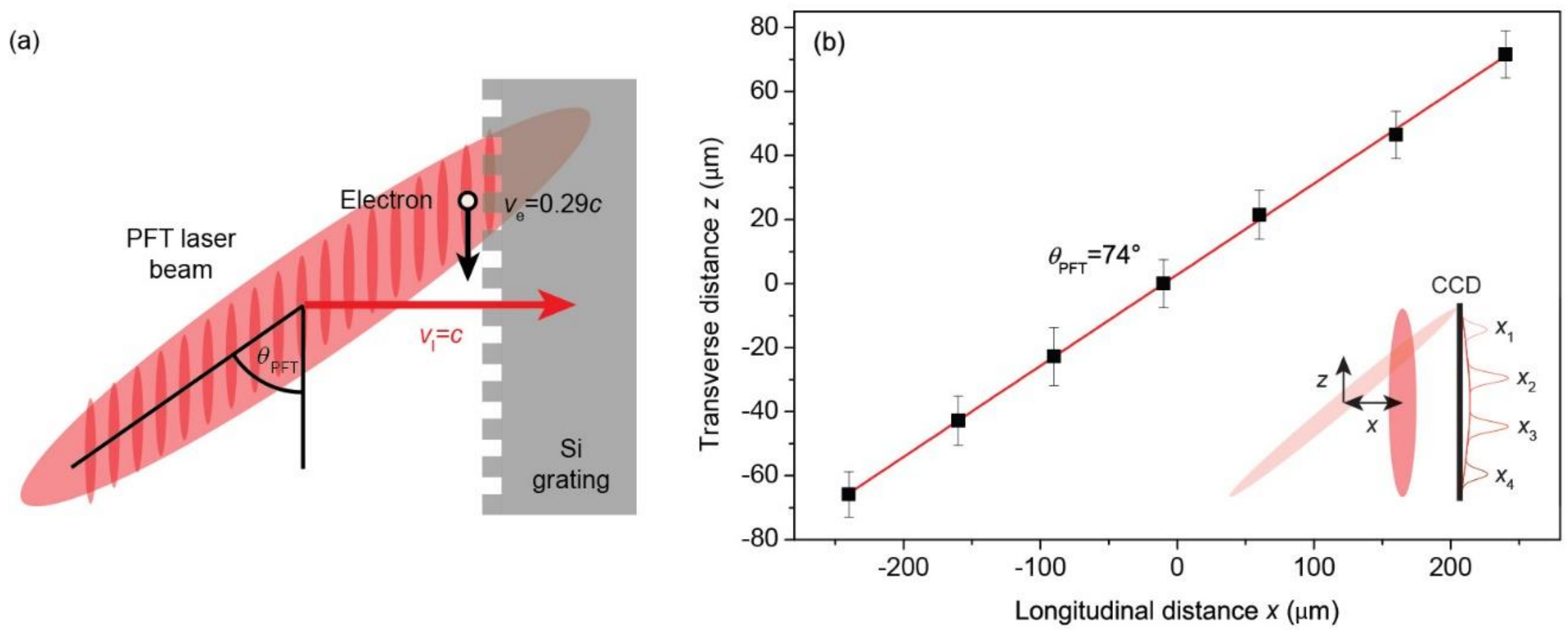

Figure 2. (a) Layout of the interaction between an electron propagating downwards along the surface of a silicon grating and optical near-fields generated by a pulse-front-tilted IR laser beam propagating horizontally from the left to the right. (b) Measured transverse position (along electron propagation direction *z*) of the peak of the cross-correlation signal between PFT and flat-intensity front laser pulses on the CCD chip as a function of the longitudinal distance between the two pulses (*x*-coordinate in the inset) adjusted by a translation stage. Inset: Sketch of the PFT characterization, where the change in the relative longitudinal distance between the PFT and flat-intensity front pulses leads to the transverse position dependence of the nonlinear detection signal on the CCD chip, labeled as $x_i$.

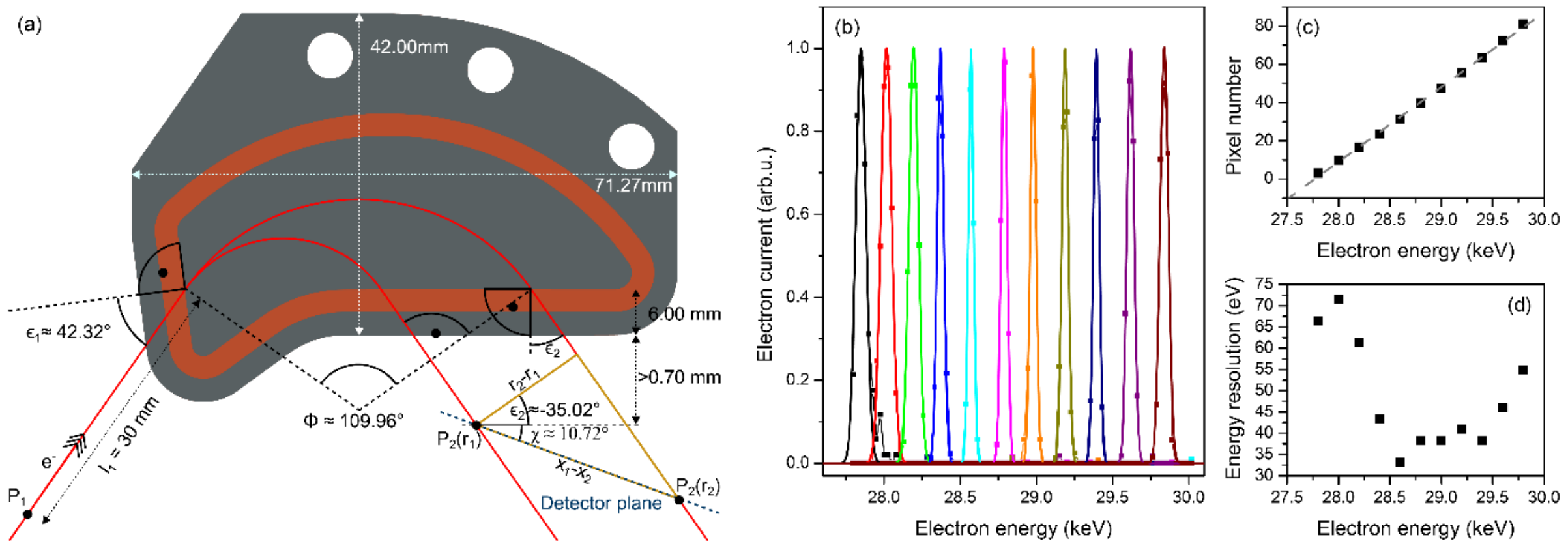

Figure 3. (a) Layout of the magnetic spectrometer. The point $P_1$ indicates the position of the input slit (focus of the electron beam). (b) Measured response functions (squares) of the magnetic spectrometer at different electron energies fitted by Gaussian functions (curves). (c) Measured dispersion curve of the spectrometer. (d) Resolution the spectrometer determined from the FWHM of the measured response functions shown in (b).

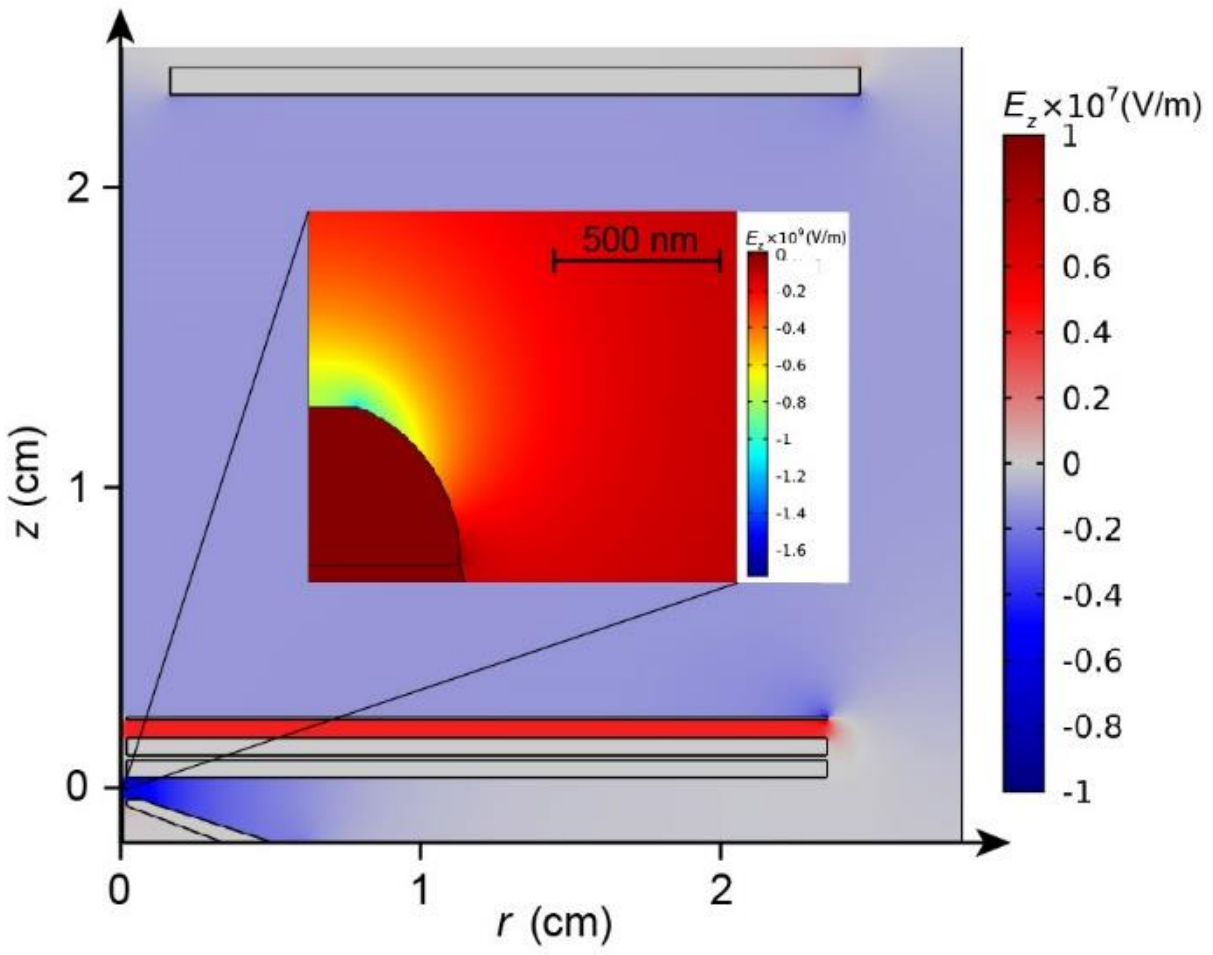

Figure 4. Geometry of the numerical model of the electron gun with the longitudinal component of the static electric field $E_z$ calculated by COMSOL multiphysics. Displayed electrodes are suppressor cylinder (lower left corner), extractor and condenser plates and grounded plate for electron acceleration to the final energy (top). Inset: Detail of the distribution of $E_z$ around the tip apex.

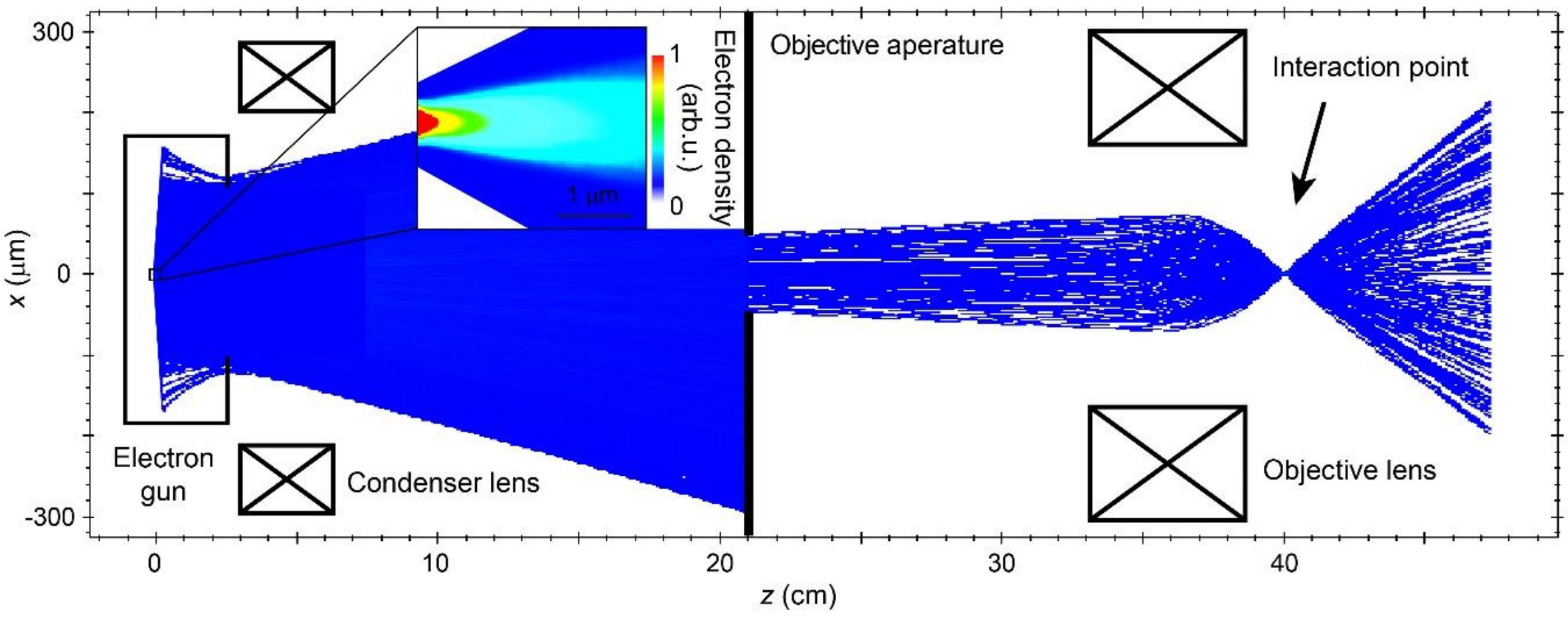

Figure 5. Electron trajectories calculated by GPT (blue lines). Inset shows the detail of the electron density close to the apex of the Schottky tip (color scale).

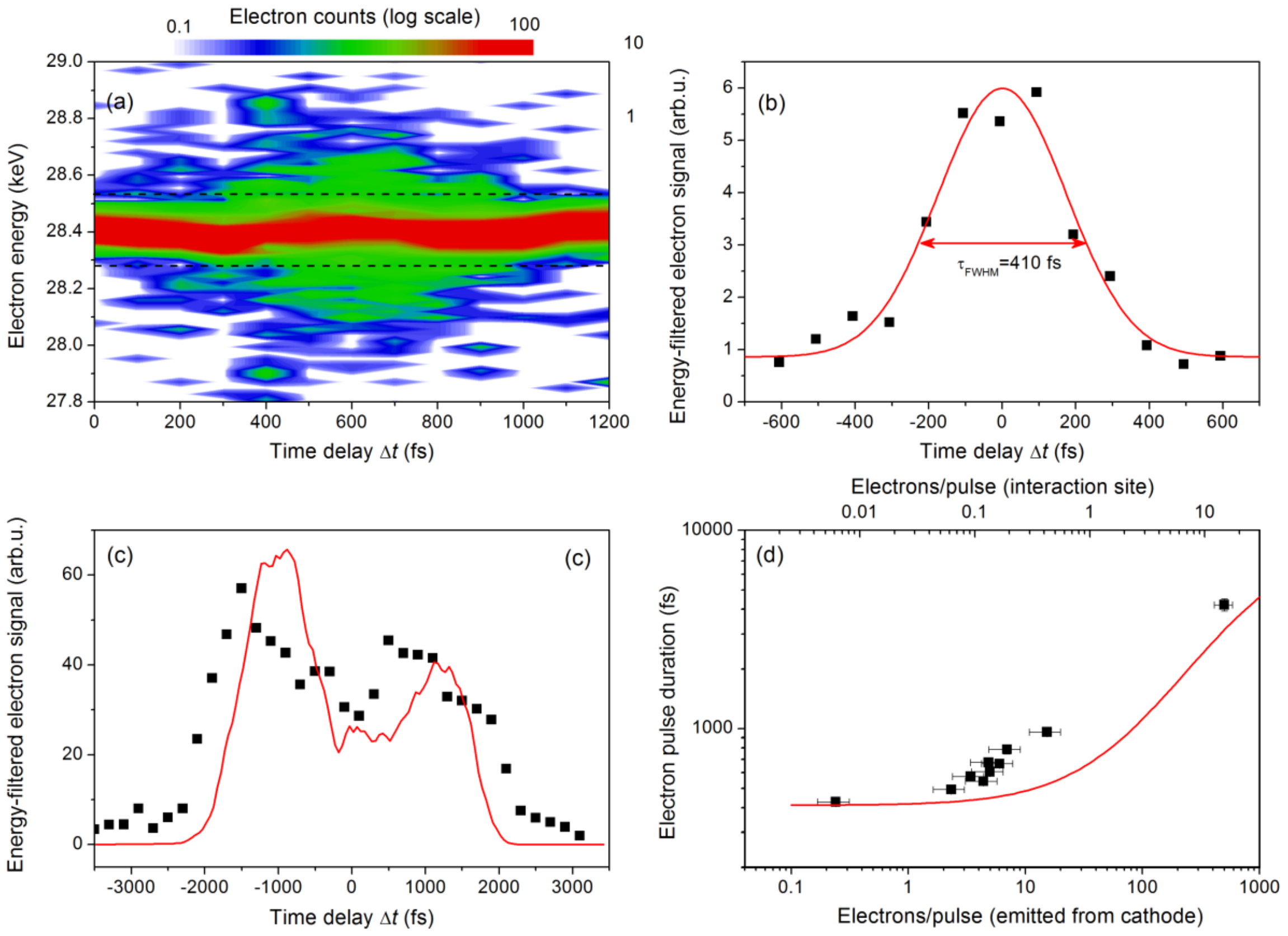

Fig. 6. Electron pulse duration measurements. (a) Electron spectra as a function of the time delay between the electron and laser pulses (in Fig. 1, $\Delta t$ is varied). (b) Cross-correlation signal obtained from (a) by integrating electrons in each spectrum out of the region of the initial electron energy spectrum marked by the two dashed lines (squares). The data is fitted by a Gaussian function with $\tau_{FWHM}$=410 fs (red curve) with an average charge per pulse emitted from the cathode ~0.2 electrons (0.03 aC). (c) Same as (b), with an average charge per pulse emitted from the cathode of 500 electrons (80 aC). Red curve represents numerical simulation including Coulomb interaction (d) Measured (squares) and calculated (red curve) electron pulse duration as a function of number of electrons emitted from the cathode per laser pulse (bottom scale) and number of electrons per pulse in the interaction region (upper scale, USEM objective lens aperture with the diameter of 100 μm).

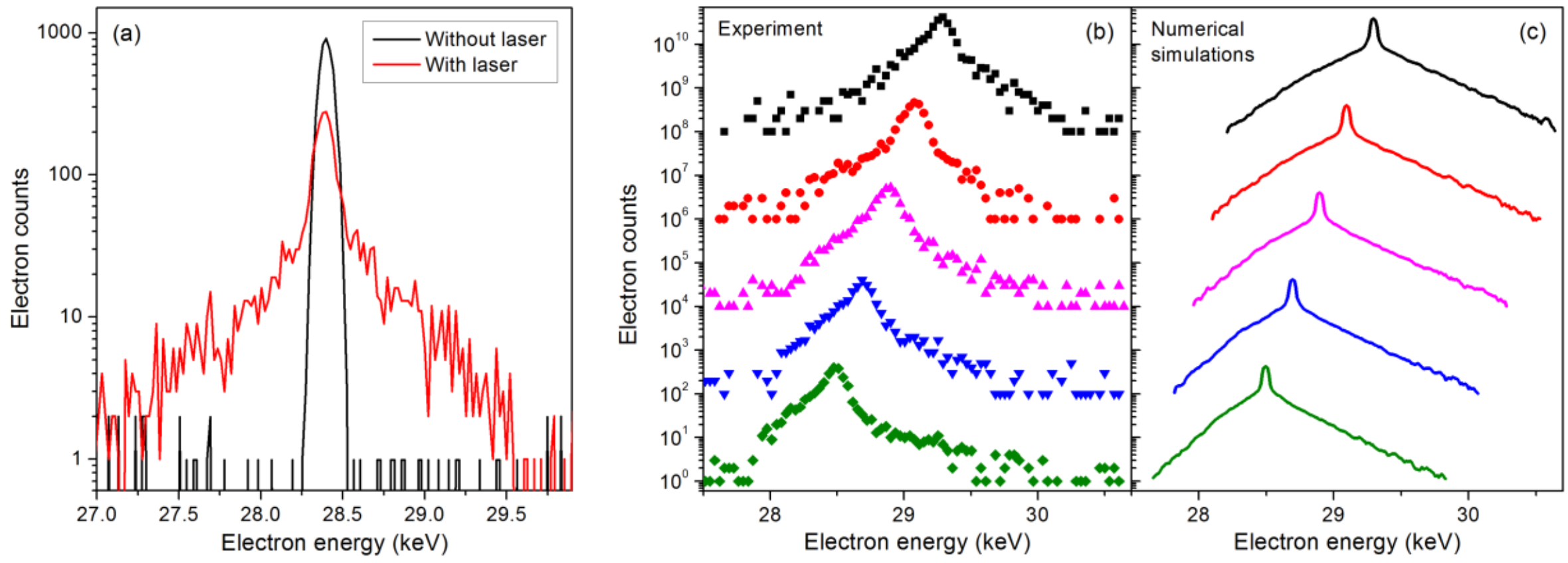

Fig. 7. (a) Electron spectra after passing by the surface of the silicon grating without (black) and with (red) the pulsed laser beam present, exciting the synchronous near-fields. (b) Measured and (c) calculated post-interaction electron spectra for different initial electron energies 28.5-29.3 keV (spectra were vertically shifted for clarity). The resonance condition is met for the black spectra. An initial electron velocity smaller than the synchronous mode phase velocity leads to an asymmetric shape of the spectra as a consequence of different dephasing lengths for accelerated and decelerated electrons.

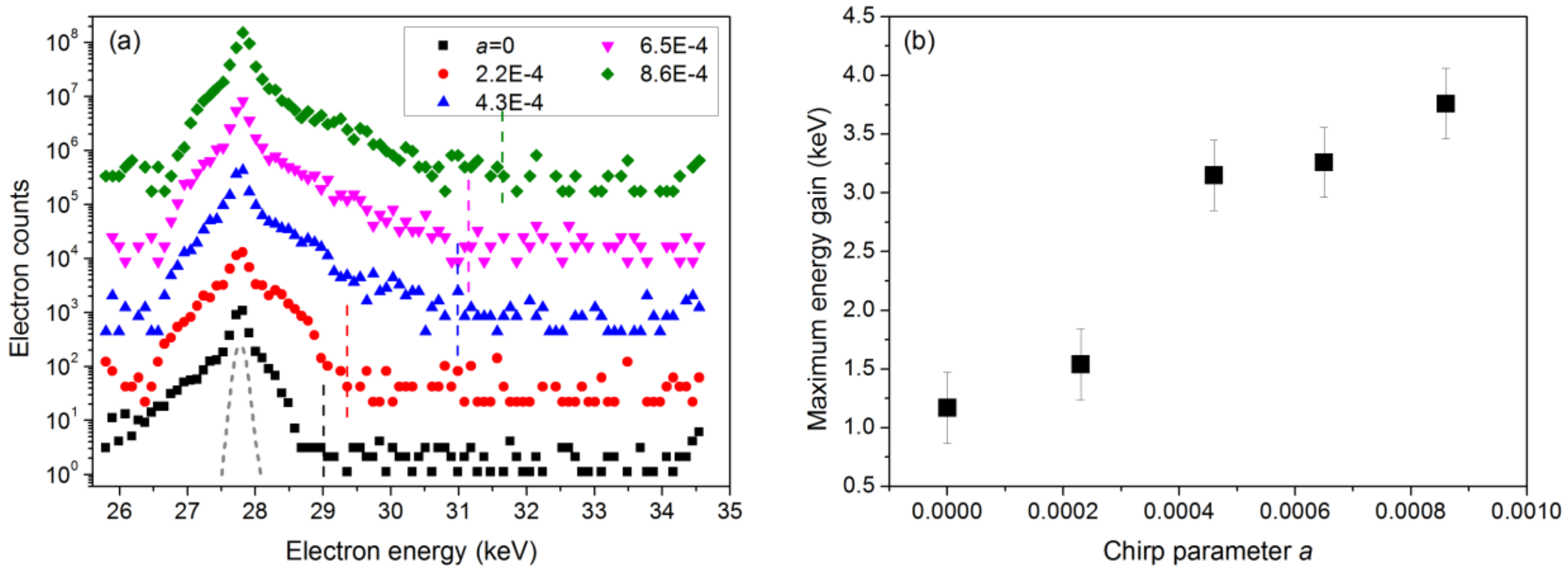

Figure 8. (a) Measured electron spectra after the interaction with optical near-fields of a linearly chirped grating structure (grating period adiabatically grows as $\lambda_p = \lambda_{p0} + az$) with different values of the linear chirp parameter $a$. Data are vertically shifted for clarity. (b) Maximum observed energy gain of the electrons as a function of $a$.